\documentclass[]{IEEEtran}
\usepackage{graphicx}
\usepackage{amsmath}
\usepackage{amssymb}
\usepackage{cite}
\begin{document}
\title{On Finding Sub-optimum Signature Matrices for Overloaded CDMA Systems}

\author{M.~Heidari~Khoozani, ~F.~Marvasti~, ~E. Azghani, M.~ Ghassemian\\
Advanced Communication Research Institute, Department of Electrical Engineering\\
Sharif University of Technology, Tehran, Iran\\
mheidari@ee.sharif.edu, marvasti@sharif.edu, azghani@ee.sharif.edu, m.ghassemian@gre.ac.uk}


\maketitle

\IEEEpeerreviewmaketitle
\begin{abstract}
The objective of this paper is to design optimal signature matrices for binary inputs. For the determination of such optimal codes, we need certain measures as objective functions. The sum-channel capacity and Bit Error Rate (BER) measures are typical methods for the evaluation of signature matrices. In this paper, in addition to these measures, we use distance criteria to evaluate the optimality of signature matrices. The Genetic Algorithm (GA) and Particle Swarm Optimization (PSO) are used to search the optimum signature matrices based on these three measures (Sum channel capacity, BER and Distance).
Since the GA and PSO algorithms become computationally expensive for large signature matrices, we propose suboptimal large signature matrices that can be derived from small sub-optimal matrices.

%
 

\end{abstract}
\section{Introduction}\label{sec:intro}
\IEEEPARstart{C}{ode} Division Multiple Access (CDMA) is a method for reliable data communication among multiple users and is the standard 3G wireless systems. The general model of a CDMA system is defined as:
\begin{equation}\label{equ:model}
Y={\bf A}X+N
\end{equation}
where ${\bf A}$ is the $m \times n$ signature matrix, $m$ is the length of signatures and $n$ is the number of users. $X$ is an $n\times 1$ user column vector and $N$ is the Additive White Gaussian Noise (AWGN) vector $N=[N_1,\ldots ,N_m]^T$, such that $N_i$'s are i.i.d random variables. For binary input CDMA system, entries of $X$ are binary (\textit{i.e.}, $\{\pm 1\}$), with uniform distribution. 

Due to the bandwidth constraint of mobile communication systems it is desirable to use more user than possible. Thus, an overloaded CDMA comes to life when the number of users exceeds the length of the signatures($m> n$); in this situation orthogonal signature matrices, such as Hadamard codes, can no longer be used.

An loading factor for such systems is defined as follows:
\begin{equation}
\beta = \frac{n}{m}
\end{equation}

Most of the work in the evaluation of sum capacity has been done for large scale CDMA systems (asymptotic results) \cite{tanaka}-\cite{korada}. 
For finite scale systems the actual capacity is not known, however there are lower and upper bounds for the sum capacity \cite{PedramIT}-\cite{Enigma}. A review of these papers is done by Hosseini in \cite{Review}.

Pad \textit{et al.}\cite{PedramIT} presented optimum signature sets for binary input and CDMA matrices such as Codes for Overloaded Wireless (COW) matrices. Furthermore, a new ML decoder is introduced for large scale COW matrices. Alishahi \textit{et al.} \cite{KasraIT} evaluated the upper and lower bounds of sum channel capacity for binary CDMA systems with and without noise. Alishahi \textit{et al.}  \cite{Enigma} generalized these previous works \cite{PedramIT} and \cite{KasraIT} to finite none binary descrete input and matrix entries. In \cite{ourpaper} we got some partial results for optimal real matrices base on GA.


In the present paper we aim to derive optimum real or binary signature matrices with binary input. For the optimality we use different measures such as sum channel capacity, bit error rate BER and in addition to them we use a distance criteria to reduce the computational complexity. It is noted that we can use these criteria for non binary input CDMA systems.

Our paper includes the following main contributions:
Firstly, we propose three main criteria for optimizing signature matrices; the channel capacity maximization, BER minimization and distance criteria. Secondly, Since it is difficult to evaluate these three criteria for large scale systems, we present a method to derive sub-optimum signature matrices by enlarging low dimensional ones.


The rest of this paper is organized as follows: 
Optimization measures and methods for optimum signature matrices derivation are discussed in Section \ref{sec:criteria}. In Section \ref{sec:optimizationtech}, numerical and simulation results based on GA and PSO are presented and compared 
Section \ref{sec:enlarging} proposes a method to derive sub-optimum high dimensional signature matrices. Finally, Section \ref{sec:conclusion} concludes the paper and highlights the future works.

\section{Signature Matrix Optimization Measures}\label{sec:criteria}

In this section we will discuss three measures for designing sub optimal signature matrices. We will show that there is a trade of between accuracy and computational complexity of these criteria. Each criterion has certain properties which is useful in specific conditions such as low SNR values and high overloading factors.

\subsection{Channel Capacity Criterion}\label{subsec:cap}
The most precise method to derive optimized signature matrices is by using the sum capacity criterion.
The sum capacity for a given dimension $n\times m$ and noise with variance $\sigma_N$ can be defined as:
\begin{equation}\label{equ:cap}
C(n,m,\sigma_N)=\max_{{\bf A}\in \mathbf{R}_{m\times n}}{C(n,m,\sigma_N|{\bf A})}
\end{equation}
where $C(n,m,\sigma_N|{\bf A})$ is the sum capacity for a specific matrix $\bf A$.
Considering that ${\bf A}$ is deterministic in (\ref{equ:model}) and (\ref{equ:cap}), we have:
\begin{equation}\label{equ:info}
C(n,m,\sigma_N|{\bf A})=\max_{P(X)}I(X;Y)=\max_{P(X)}h(Y)-h(N)
\end{equation}
where $h(y)$ is the differential entropy. According to the conjecture mentioned in \cite{Shayan} and \cite{KasraIT}, $h(y)$ is maximized when $X$ is uniform. For input uniform distribution since $N_i$'s are \textit{i.i.d.} and thus $f_N(N)=\prod_{i=1}^m{f_{n_i}(n_i)}$, the probability distributed function (pdf) of $Y$ is as bellow:
\begin{align}\label{equ:fy}
f_{Y}&(Y ) =\frac{1}{2^n} \times\nonumber\\
&\sum_{\bar{X}\in \{\pm 1\}^{n\times 1}}{\Bigg[\Big(\frac{1}{2\pi \sigma_N^2 }\Big)^{\frac{m}{2}}\prod_{i=1}^m \exp\left(\frac{-\left(y_i-{\bf A}_{i} \cdotp {X}\right)^2}{2\sigma_N^2}\right)\Bigg]}
\end{align}
where $y_i$ and ${\bf A}_{i}$ are the $i^{th}$ entry of $Y$ and the $i^{th}$ row of matrix ${\bf A}$, respectively. As a result of this, $h(Y)$ is computable from (\ref{equ:fy}) as:
\begin{equation*}\label{equ:h_y}
h_Y(Y)=\underbrace{\int \ldots \int}_m f_Y(Y)\log_2(f_Y(Y ))\mathrm{d}y_1\ldots\mathrm{d}y_m
\end{equation*}

 Consequently $C(n,m,\sigma_N|{\bf A})$ is derived from the mutual entropy in (\ref{equ:info}). The per-user sum capacity parameter is used instead of channel capacity, which can be defined as:
\begin{equation}
c=\frac{C_{Channel}}{n}
\end{equation}
where $c$ is the normalized of $C$.
It must be mentioned that the computational complexity of capacity measure is of $O(m^n)$ which results in an NP-hard algorithm.

\subsection{BER Criteria}\label{subsec:BER}
To compute the BER measure, a large array of bits ($10^6$ bits) are produced, encoded, transmitted through the simulated channel and decoded at the receiver by an ML decoder. After the decoder, the probability of error statistically measured. We conjecture that the lower the BER of a signature matrix is, the higher its channel capacity will be; this conjecture is verified with simulation results. Hence, optimum matrices can be derived by minimizing the BER measure. 

The BER requires much less computation than the capacity evaluation. However both criteria are not practical due to extensive computation.  Therefore, we propose distance measures as discussed bellow.

\subsection{Distance criterion}\label{subsec:constellation}
In this subsection, we propose three different methods based on the output
constellation points ($Z_{i}$), \textit{i.e.}, The output points which are defined in the absence of noise:
\begin{equation}
Z_{i}={\bf A}X_{i}\quad X_1,X_2,\cdots,X_{2^n}\in\{ \pm 1 \}^{n\times 1}
\end{equation} 

\begin{figure}[h]
\centering
\includegraphics[width=9cm]{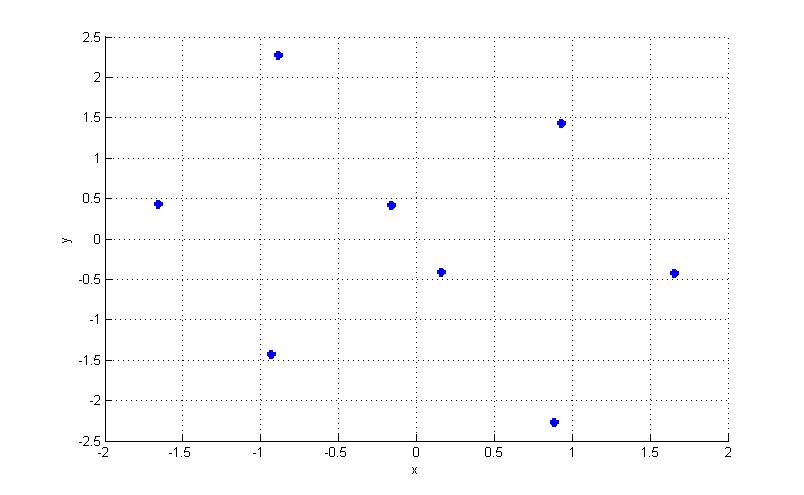}
\caption{Constellation of a $2\times 3$ signature matrix optimized for $E_b/N_0=5dB$ using the channel capacity criteria}
\label{fig:Constellation}
\end{figure}
Fig. \ref{fig:Constellation} shows an example for constellation of a $2\times 3$ real signature matrix (which is optimized using channel capacity method for $E_b/N_0=5dB$). This constellation is the projection of $\pm 1$ three dimensional cube on a plane.
Since $X_i, -X_i \in \{\pm 1\}^{n\times 1}$ for every $i=1,2,\cdots 2^n$ we have a similar symmetry property for $Z_i$'s. Therefore, the computational complexity is reduced by considering only half of $Z_i$'s in the distance criterion.
One way for optimization is maximizing the minimum distance of the output constellation points. This method guarantees an upper bound for the probability of error for high values of SNR.
The following equation explicitly defines \textit{Minimum Distance} (MD) criterion:
\begin{equation*}\label{equ:mindis}
{\bf MD} =\min_{i \neq j}\ ||Z_i-Z_j||
\end{equation*}
where $||U||$ represents the Euclidean norm of the vector $U$. Note that MD must be maximized in order to find a sub-optimum matrix. 
By taking a more analytical approach, suppose that $x_i$'s (the $i^{th}$ element of the input vector $X \in \{ \pm 1 \}^{n\times 1}$) are independent, hence the error probability of a block (with size of $m$) is derived as follows:
\begin{equation}\label{equ:P_e}
P_e=\frac{1}{2^n}\sum_{i=1}^{2^n}{P\bigg{(}\bigcup_{\begin{smallmatrix} k=1 \\ k\neq i \end{smallmatrix}}^{2^n}||Y_i-Z_i||^2 > ||Y_i- Z_k||^2\bigg{)}}
\end{equation}
where $Y_i=Z_i+N$ and $Z_i$ is the output vectors in the noiseless channel. Since the noise vector elements are \textit{i.i.d.} Gaussian random variables with variance $\sigma_N$, the upper bound for error probability can be calculated as:
\begin{equation}\label{equ:se7en}
P_e < \frac{1}{2^n}\sum_{i=1}^{2^n}{\sum_{\begin{smallmatrix} j=1 \\ j\neq i \end{smallmatrix}}^{2^n}Q\left(\frac{||Z_i-Z_j||}{2\sigma_N}\right)}
\end{equation}
where $Q(x)$ refers to Cumulative Distribution Function (CDF) of Gaussian distribution.
Instead of maximizing MD, we minimize the upper bound given in (\ref{equ:se7en}) which is equivalent to  minimizing the following function:  
\begin{equation}\label{equ:QD}
QD =\ \sum_{i=1}^{2^n}{\sum_{\begin{smallmatrix} j=1 \\ j\neq i \end{smallmatrix}}^{2^n}Q\left(\frac{||Z_i-Z_j||}{2\sigma}\right)}
\end{equation}
The minimization of the upper bound can give the better results than MD; however it is more computationally intensive. To reduce the computational complexity of $Q(x)$, the following approximation can be used the following approximation:
\begin{equation}\label{equ:curvfit}
Q(x)\ \simeq\ 0.7 \exp{\left(-\left(\frac{x+1}{1.6}\right)^2\right)}
\end{equation}
Thus we can use the following distance denoted by ED:
\begin{equation}\label{equ:ExpDist}
ED =\ \sum_{i=1}^{2^n}{\sum_{\begin{smallmatrix} j=1 \\ j\neq i \end{smallmatrix}}^{2^n}\exp\Bigg[-\left(\frac{\frac{||Z_i-Z_j||}{2\sigma_N}+1}{1.6}\right)^2\Bigg]}
\end{equation}
For high SNR values, (\ref{equ:ExpDist}) is simplified to a single exponential element which presents a similar behaviour for MD. The next section is a brief description for GA and PSO. 

\section{Optimization Techniques}\label{sec:optimizationtech}
In this section, we will describe the optimization techniques to find good signature matrices based on the criteria discussed on Section \ref{sec:criteria} that are applied to optimize the criteria. We use two known optimization techniques, namely GA and PSO.
We show that since GA and PSO are methods for minimizing an arbitrary cost function, treating the criteria as cost functions, will lead to derive optimum signature matrices. 

\subsection{Genetic Algorithm}
The Genetic Algorithm (GA) \cite{GA} employs the principal of survival of the fittest in its search process to select and generate individuals (design solutions) that are adapted to their environment (design objectives/constraints). Therefore, over a number of generations (iterations), desirable traits (design characteristics) will evolve and remain in the genome composition of the population over traits with weaker undesirable characteristics. The GA is well suited and has been extensively applied to solve complex design optimization problems because it can handle both discrete and continuous variables with nonlinear objective and constraint functions. In this work we apply the GA to find sub optimum signature matrices based on criteria of sum capacity, BER, and the distance. The next algorithm is an alternative algorithm that can converge to some optimal matrices with the faster rate of convergence. The parameters and options of the GA are presented in \mbox{Table \ref{tab:GA}}.
\begin{table}[h]
\caption{Simulation parameters set for GA}
\begin{tabular}{l p{4cm}}
\large{Population}& \\\hline
Size & 20 individual matrices \\ 
Type & Double vector \\
Creation function & Uniform in first run
Best results of previous runs afterwards \\
Lower bound &-1 \\ 
Upper bound & +1 \\
\begin{scriptsize}
•
\end{scriptsize}\\
\large{New Generation}\\\hline
Elite count & 2\\
Crossover fraction &0.8\\
Migration direction & Forward\\
Mitigation factor & 0.2 \\
\\
\large{Stopping criteria}\\\hline
Number of iterations & 100\\
Function tolerance & $10^{-6}$\\
\end{tabular}
\label{tab:GA}
\end{table}
\subsection{Particle Swarm Optimization}
PSO \cite{PSO1},\cite{PSO2} similar to GA is a computational method that optimizes a problem by iteratively trying to improve a candidate solution, which results in an objective function. In PSO, a set of randomly generated solutions (initial swarm) propagates in the design space towards the optimal solution over a number of iterations (moves) based on large amount of information about the design space that is assimilated and shared by all members of the swarm.
The PSO algorithm considers some candidate solutions in the search domain. During each iteration, the cost function of each candidate solution is calculated. Each candidate solution can be considered as a particle moving toward the minimum value of the cost function. As the first step, PSO chooses the candidate solutions randomly inside the search space. It should be mentioned that the PSO does not have any prior information about the cost function; it does not know which particles are near or far from the global minimum of the cost function. What PSO does, is to evaluate the cost value of each particle and just work with the corresponding cost values. The position of a particle is composed of its candidate solution, cost and velocity. Moreover, it remembers the least cost (the best fitness) that it has had thus far during the operation of the algorithm, called the individual best fitness. The candidate solution corresponding to this fitness is referred to as the individual best candidate solution or the individual best position. At last, the PSO seeks for and finds the least cost among all the particles in the swarm, named the global best fitness. Bellow we will compare the GA and PSO in terms of convergence rate. The simulation parameters and options set for the PSO algorithm are listed in Table \ref{tab:PSO}.
 \begin{table}[h]
\caption{Simulation parameter set for PSO}
\begin{tabular}{l p{4cm}}
\large{Particle’s position}\\ \hline
Number of particles & 20 individual vectors\\
Initial position & Random, uniform distribution\\
Lower and upper bounds & [−1, +1]\\
Particle Velocity & initiated randomly with uniform distribution\\
\\
\large{Stopping criteria}\\\hline
Number of iterations & 100\\
\end{tabular}
\label{tab:PSO}
\end{table}

\subsection{ Convergence evaluation of GA and PSO}
\begin{figure}[h]
\centering
\includegraphics[width=9cm]{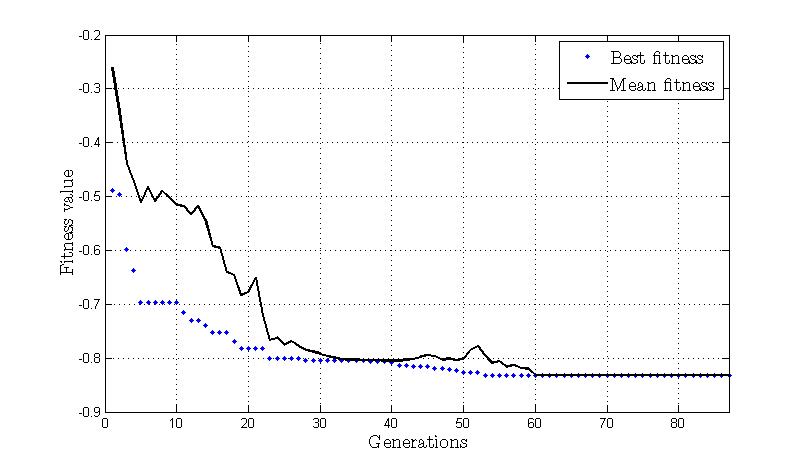}
\caption{The best value and the mean value of populations versus GA iterations for the MD method ($\beta=\frac{5}{4}$).}
\label{fig:Genetic}
\end{figure}

\begin{figure}[h]
\centering
\includegraphics[width=9cm]{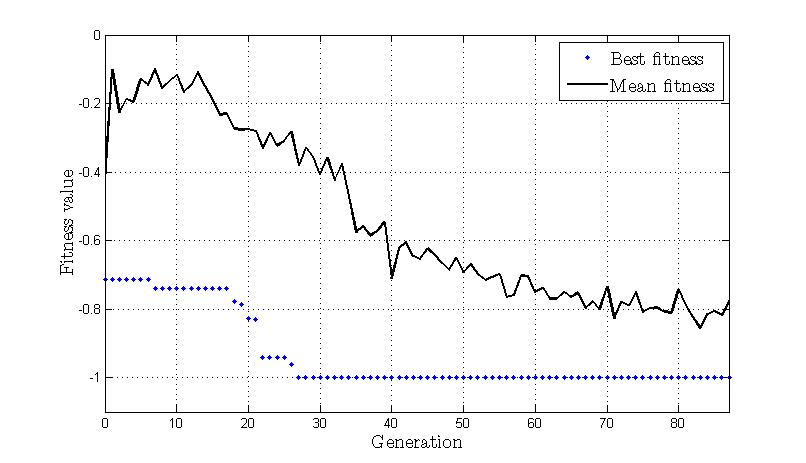}
\caption{The best value and the mean value of populations versus PSO iterations for the MD method ($\beta=\frac{5}{4}$).}
\label{fig:PSO}
\end{figure}

We check the convergence behavior of the discussed optimization algorithms before we apply them for our analysis. It is important for GA to converge to the minimum point for each method. We can show the convergence rate by comparing, the best and the mean fitness values for every iteration that are the minimum and the mean of objective function in each iteration, respectively. For instance, considering MD as an objective function for $n = 5$ and $m = 4$, there is no difference between the best and the mean values after $50$ iterations (as shown in Fig. \ref{fig:Genetic}), therefore, one concludes convergence of the GA . Furthermore, the variation of the mean fitnesses in various iterations shows that the GA explores almost the whole of the feasible populations, which implies global minimum as oppose local minimum. 
The best and the mean fitness values for each iterations for the PSO algorithm is depicted in Fig.~\ref{fig:PSO}, using MD as the objective function when $n = 5$ and $m = 4$. In this figure unlike Fig. \ref{fig:Genetic} the beast and the mean do not cross each other, however the best value is unchanged after a certain iteration. This figure shows that the algorithm converges after $28$ iterations. A comparison between Fig.~\ref{fig:Genetic} and Fig.~\ref{fig:PSO} demonstrates that the PSO converges faster than the GA. 

%
%
\section{Numerical and Simulation Results}\label{subsec:Simulation}
This section presents the numerical and simulation results of the GA and PSO for proposed criteria, and compare GA and PSO in term of optimality. Sub-optimum signature matrices based on various criteria using GA and PSO are presented in the Appendix.
\subsection{Results of GA for Real Valued Signature  Matrices}

In this subsection, we apply the GA to verify our results discussed in Section \ref{sec:criteria}, namely BER, channel capacity and distance methods. Firstly, we will compare distance methods with each other and then, we will show the simulation and numerical results of various methods.


\subsubsection{\bf Comparison of Distance Criteria}\label{subsec:Constellation factors}
\begin{figure}[h]
\centering
\includegraphics[width=9cm]{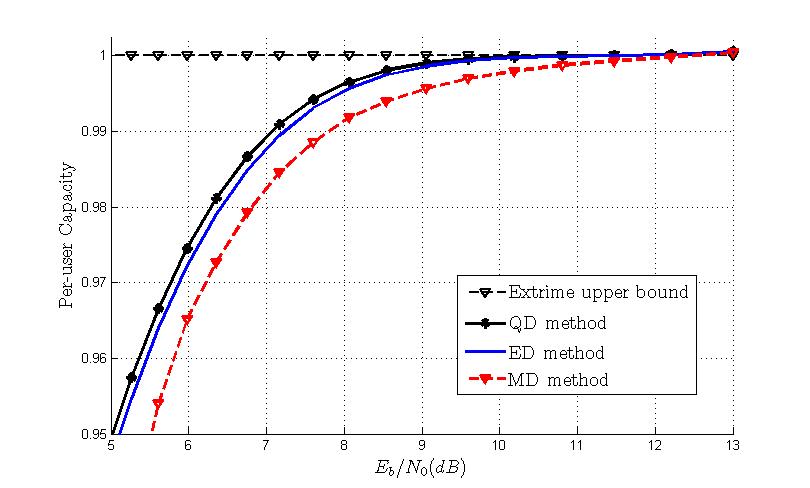}
\caption{Per-user capacity of sub-optimum matrices based on distance methods versus different $E_b/N_0$ values ($\beta=\frac{4}{3}$, using the GA)}
\label{fig:C45}
\end{figure}

In order to compare distance criteria in term of capacity. We obtain sub-optimal matrices for an arbitrary $E_b/N_0$ value and compare the capacity of these matrices for that $E_b/N_0$ value. Fig. \ref{fig:C45} illustrates the per-user capacity of optimized matrices using distance criteria for different $E_b/N_0$ values.
The matrices exhibit a per-user channel capacity curve close to the extreme upper bound, i.e., 1 bit per second per user.
The minor difference between the curve of ED and QD method justifies the accurate approximation of QD in (\ref{equ:QD}) and (\ref{equ:curvfit}). As is expected in Subsection \ref{subsec:constellation}, the MD criteria capacity performance is close to ED and QD methods for high $E_b/N_0$ values, which is shown in Fig. \ref{fig:C45}. Due to the fact that the ED method is closed to QD but with less complexity, we will choose the ED as the distance measure to compare with other criteria such as BER and sum capacity criteria.

\subsubsection{\bf Comparison with Criteria}

\begin{figure}[h]
\centering
\includegraphics[width=9cm]{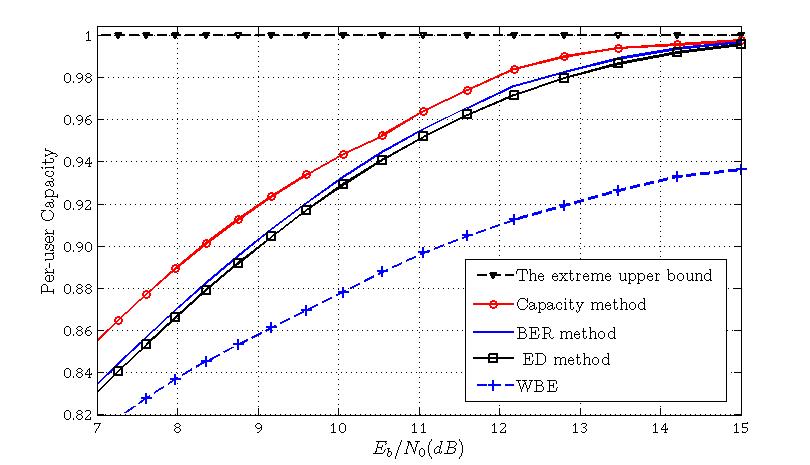}
\caption{Per-user capacity of sub-optimum matrices based on different criteria for various $E_b/N_0$ values($\beta = \frac{5}{2}$, using the GA)}
\label{fig:comparison}
\end{figure}

\begin{figure}[h]
\centering
\includegraphics[width=9cm]{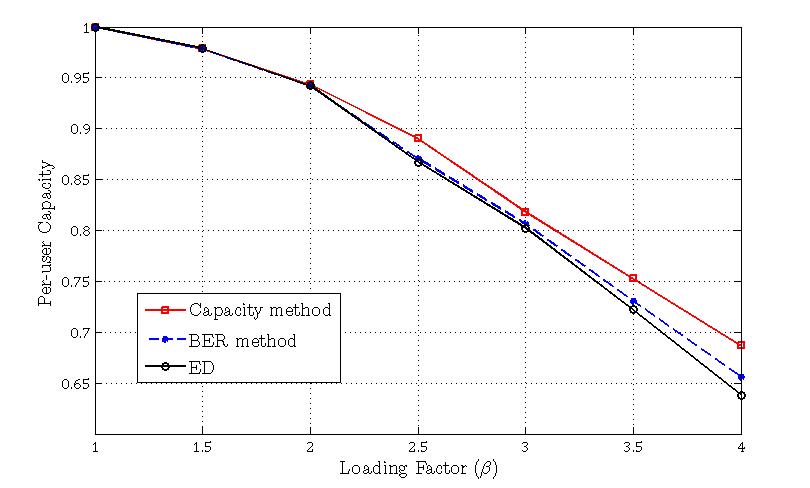}
\caption{Per-user capacity versus loading factor ($E_b/N_0=8dB$, using the GA).}
\label{fig:overload}
\end{figure}

Fig. \ref{fig:comparison} demonstrates the per-user capacity of different proposed matrices optimized by BER, capacity, and ED criteria for different $E_b/N_0$ values. In addition, we compare these results with the Welch Bound Equality (WBE) codes which are introduced in \cite{Welch} and \cite{Welch2}. Since WBE is optimum for Gaussian input distribution, there is no guarantee to be optimum for binary input vectors. Among the proposed matrices, the results of the ED method are close to that of BER scenario; this verifies that our approximation in (\ref{equ:se7en}) and (\ref{equ:ExpDist}) is accurate. As discussed earlier, to compute the BER criterion, a large array of bits needs to be processed which makes the computation of this criterion more complex compared to the ED criterion.
Fig. \ref{fig:overload} shows the per-user capacity of optimized matrices based on the capacity method, ED, and BER method versus loading factor ($\beta=\frac{n}{m}$). in this figure we fixed an $E_b/N_0$ value ($8dB$) to compute the sum capacity. The performance of the capacity method decrease slower than other methods which shown the robustness of this method against loading factor. 
\subsection{ Results of GA for Binary Valued Signature Matrices}
\begin{figure}[h]
\centering
\includegraphics[width=9cm]{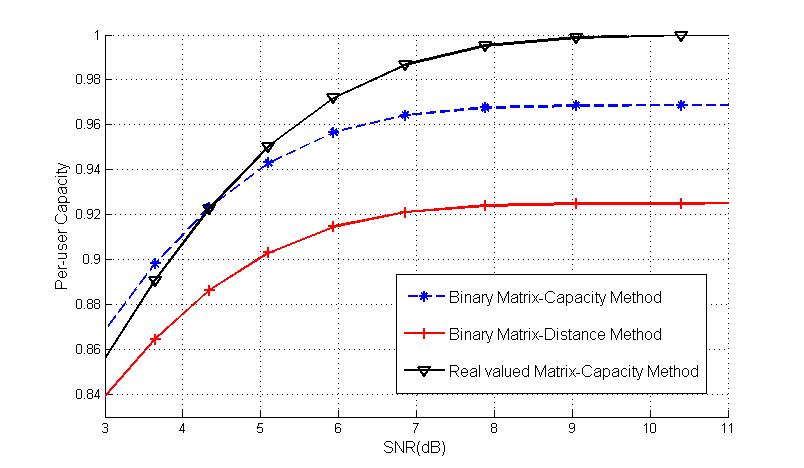}
\caption{Per-user capacity of binary matrices and real valued matrices ($\beta=\frac{5}{4}$, using the GA).}
\label{fig:binary}
\end{figure}

Although, our proposed methods are applied to real valued matrices, we can use them to find sub optimal binary ($\pm 1$) matrices.
Binary matrices are much simpler, in implementation, than real valued ones. We consider an $4\times 5$ binary matrix ($A_5$ at Table \ref{tab:GA_CAP}) which is optimized by the capacity criterion  and compare it with another binary matrix derived from the ED criterion ($A_3$ at Table \ref{tab:GA_ED}).
Fig. \ref{fig:binary} shows per-user capacity of these two binary matrices with a real valued sub-optimum matrix derived from the sum capacity criterion ($A_4$ at \ref{tab:GA_CAP}). This figure shows that the binary matrices developed by sum capacity criterion can be close to the real valued matrices however the binary matrix derived from the ED method is not as good.
\subsection{Results of PSO}\label{subsec:PSO}
In this subsection, we present the simulation and numerical results of the PSO. Due to poor result of this algorithm as appos to GA for binary matrices, we only show the results for real valued signature matrices.
\begin{figure}[h]
\centering
\includegraphics[width=9cm]{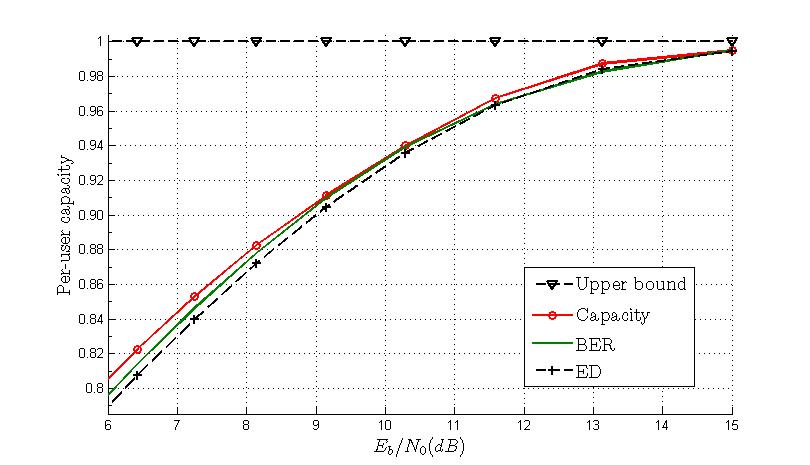}
\caption{Per-user capacity of sub-optimum matrices based on different criteria for various $E_b/N_0$ values ($\beta=\frac{5}{2}$, using the PSO).}
\label{fig:heavyPSO}
\end{figure}

\mbox{Fig. \ref{fig:heavyPSO}} shows the performance of the sub-optimized matrices derived by PSO for the case when $\beta=\frac{5}{2}$. The curves of the BER, MD, and the capacity criteria are near the upper bound (1 bit per sec per user). In this figure, the results of the BER and the ED methods are very close to the capacity method unlike the GA depicted in Fig. \ref{fig:comparison}.

\begin{figure}[h]
\centering
\includegraphics[width=9cm]{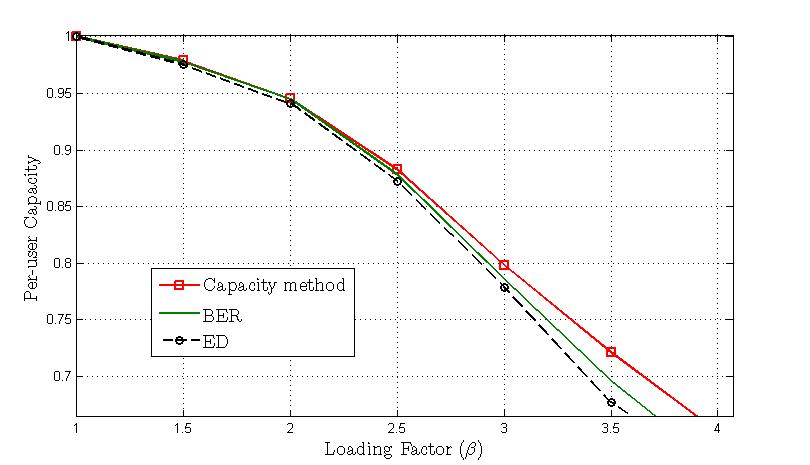}
\caption{Per-user capacity of sub-optimum matrices versus loading factor ($E_b/N_0=8dB$, using the GA).}
\label{fig:PSO-overload}
\end{figure}

Fig. \ref{fig:PSO-overload} shows the per user capacity versus the loading factor for various criteria.
As expected, the performances of all the cases decrease with increasing $\beta$. Similar to the GA the Capacity and BER criteria are the best criteria for large $\beta$.
 
\subsection{Comparison of GA and PSO Results}
\begin{figure}[h]
\centering
\includegraphics[width=9cm]{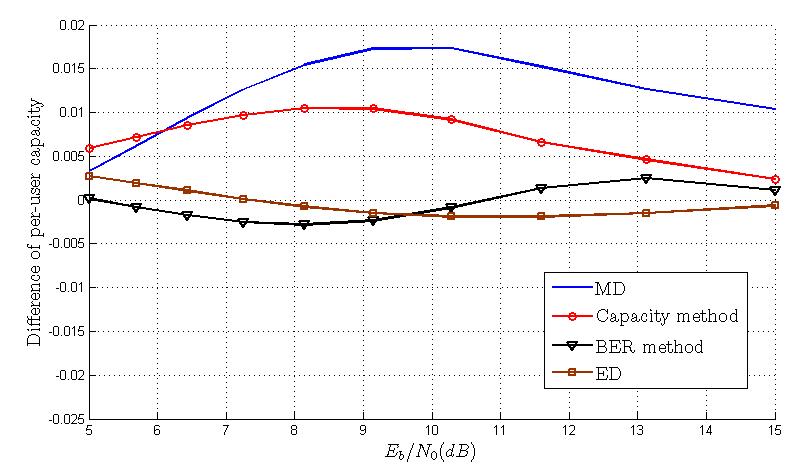}
\caption{Per-user capacity derived from the GA minus that from the PSO for various measures versus the $E_b/N_0$ ($\beta=\frac{5}{2}$).}
\label{fig:PSO-GA25}
\end{figure}

\begin{figure}[h]
\centering
\includegraphics[width=9cm]{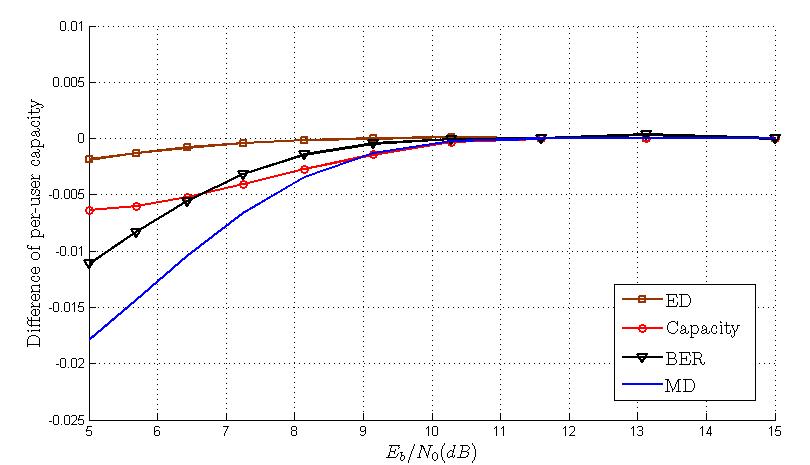}
\caption{Per-user capacity derived from the GA minus that from the PSO for various measures versus the $E_b/N_0$ ($\beta=\frac{4}{3}$).}
\label{fig:PSO-GA34}
\end{figure}

In this subsection, we evaluate the sensitivity of the proposed criteria with respect to the loading factor and optimization algorithms (GA and PSO). Fig. \ref{fig:PSO-GA25} demonstrates a comparison between the GA and the PSO for all the criteria  ($\beta=\frac{5}{2}$). This figure shows per-user capacity of GA minus the PSO algorithm  for various. Also, Fig.\ref{fig:PSO-GA34} presents the same results when $\beta=\frac{4}{3}$. A comparison between fig 10 and Fig. 11 shows that for low loading factors ($\beta$) and small values of Eb/N0 the PSO performs better than the GA. On the the hand, for high values of betta the GA performs better. Also, these two figure show that for the ED and BER criteria the choice of the GA and the PSO algorithms do not make any differences.

Note as discussed in Section II the PSO algorithm is about 5 times faster than the GA algorithm. Although note that, in general, the GA algorithm performs slightly better that PSO for real signature matrices  but it is the only choice for binary matrices.

In the following section we propose a method to design large size signature matrices for increasing the sum capacity.

\section{Designing Large Signature Matrices}\label{sec:enlarging}
The GA and the PSO algorithms are not suitable for large signature matrices such as $64\times128$. We thus propose a method of enlarging signature matrices from a sub-optimal small scale signature matrix. This section presents the method for enlargement from sub-optimal signature matrices.

 We propose to derive optimum matrices in a different way. Instead of the proposed methods discussed in Section II we derive sub-optimal signature matrices from small values of $n$ and m and then by tensor products we construct a much larger signature matrix called an enlarged matrix.
 
 Instead of directly deriving optimum matrices using the proposed methods discussed in Section \ref{sec:criteria}, we derive an optimum signature matrix for a certain small value of $n$ and $m$ and later using ``Kronecker product" yields sub-optimum matrix for the given dimension. We refer to this matrix as an ``enlarged matrix". This section presents the theoretical proof for the practicality of the proposed enlarging method as well as the proposed ML decoder design.
 In this paper we manly consider signature matrices for binary and non binary cases [part II and Part III].  Here we try to extend these results real signature matrices for any types of inputs.

\subsection{Design Procedure}
The following theorem provides the necessary mathematical ground for enlarging signature matrices.

\newtheorem{mytheorem}{\textbf{Theorem}}
\begin{mytheorem}\label{Th:1}
Suppose that $\bf{A}$ is a real $m\times n$ signature matrix for a CDMA system with binary or non binary input and $\bf{G}$ is a $k\times k$ reversible matrix such that its columns are normalized. Also, assume $C(kn,km,\sigma_N |{\bf A})$ is the channel capacity assigned to matrix $\bf A$ in the presence of AWGN noise with variance $\sigma_N$. Denote $\otimes$ as the Kronecker product and let ${\bf B}={\bf G}\otimes {\bf A}$, then,
%
\begin{equation}
C(kn,km,\sigma_N |{\bf B})\leq k{C(n,m,\sigma_N |{\bf A})}
\end{equation}
The equality holds if and only if $\bf G$ is unitary.
\end{mytheorem}

\begin{IEEEproof}
Considering (\ref{equ:model}) as the model of CDMA systems and $\bf B$ as the signature matrix, we have:
\begin{equation}\label{Y_km}
Y_{km\times 1}={\bf B}X_{kn\times 1}+N_{km\times 1}
\end{equation}
where $N$ is the noise vector with variance $\sigma_N$. For an arbitrary vector $U$, define $U_i$ as the $i^{th}$ segment of $U$ containing $m$ entries; thus \mbox{$U^T=[U_1,U_2,\dots,U_k]$}.
Multiply both sides of (\ref{Y_km}) by ${\bf D}={\bf G^{-1}} \otimes {\bf I_m}$ and define $Z={\bf D} Y$, hence:
\begin{equation*}
Z=({\bf G^{-1}} \otimes {\bf I_m})({\bf G} \otimes {\bf A})X+({\bf G^{-1}} \otimes {\bf I_m})N
\end{equation*} 
From the Kronecker product properties, we have:
\begin{eqnarray*}
({\bf G^{-1}} \otimes {\bf I_m})({\bf G} \otimes {\bf A})={\bf G^{-1}G} \otimes {\bf I_m A}={\bf I_k} \otimes {\bf A}
\end{eqnarray*}
Denote $M={\bf D}N$. Thus for each $i$, the entries of $M_i$ are independent Gaussian random vectors with variance \mbox{$\sigma_{M,i}=||{\bf G^{-1}}(i,:)||_2\sigma_N$}.
Having $h(Z)$, we can calculate $h(Y)$ as follows:
\begin{equation}\label{equ:hyhz}
h(Y)=h(Z)-\log_2{|\det{{\bf D}}|}
\end{equation}
From the Kronecker product properties:
\begin{equation}\label{equ:Kron-det}
\det{(\bf G^{-1}} \otimes {\bf I_m)}=(\det{\bf G^{-1}})^m (\det{{\bf I_m}})^k
\end{equation}
Considering (\ref{equ:hyhz}) and (\ref{equ:Kron-det}), we have:
\begin{eqnarray*}
h(Y)=h(Z)+m\log_2{|\det{\bf G}|}
\end{eqnarray*}
Similarly, for $N$ and $M$, we have:
\begin{equation}\label{equ:hN-M}
h(N)=h(M)+m\log_2{|\det{\bf G}|}
\end{equation}
Bear in mind (\ref{equ:info}), we can write, 
\begin{equation}\label{equ: C-hz}
C(kn,km,\sigma_N |{\bf B})=\max_{P(X)}h(Z)-h(M)
\end{equation}
Since $Z=[Z_1, Z_2, \cdots , Z_k]^T$, we have the following upper bound for $h(Z)$:
\begin{equation}\label{equ: hZ-hZ1}
h(Z)\leq\sum_{i=1}^{k}{h(Z_i)}
\end{equation}
Considering the relation between $Y$ and $Z$ and by using (\ref{equ:info}), we can derive the following equation:
\begin{eqnarray*}
\max_{P(X)}h(Z_i)=C(n,m,\sigma_{M,i} |{\bf A})+\frac{m}{2}\log_2{2\pi e \sigma_N^2}-m\log_2{||{\bf G^{-1}}(i,:)||}
\end{eqnarray*}
where ${\bf G^{-1}}(i,:)$ is the $i^{th}$ column of ${\bf G^{-1}}$. From (\ref{equ:info}) for $Y$, we have:
\begin{eqnarray*}
C(kn,km,\sigma_N |{\bf B})\leq \sum_{j=1}^k C(n,m,(||{\bf G^{-1}}(i,:)||\sigma_N )|{\bf A})\\+m\log_2{\frac{|\det{\bf G}|}{\prod_{i=1}^k ||{\bf G^{-1}}(i,:)||}}
\end{eqnarray*}
If $\bf G$ is unitary, the equality holds and we have:
\begin{equation*}
C(kn,km,\sigma_N |{\bf B})=k{C(n,m,\sigma_N |{\bf A})}
\end{equation*}
Assume, $\bf G$ is not unitary. Since columns of $\bf G$ are normalized,  $|\det{\bf G}|<\prod_{i=1}^k ||{\bf G}(:,i)||=1$ and $||{\bf G^{-1}}(i,:)||\geq 1$. Thus, due to the reverse relationship between the sum-capacity and noise power, $C(n,m,(||{\bf G^{-1}}(i,:)||\sigma_N )|{\bf A})\leq C(n,m,\sigma_N )|{\bf A})$ and,
\begin{equation*}
C(kn,km,\sigma_N |{\bf B})<k{C(n,m,\sigma_N |{\bf A})}
\end{equation*}
Therefore, $C(kn,km,\sigma_N |{\bf B})$ is maximized when  $\bf G$ is unitary.
\end{IEEEproof}
  
According to this theorem the best choice of $\bf G$ for enlarging a signature matrix is an unitary matrix. In practice, we use a binary Hadamard matrix with normalized columns as $\bf G$. Let $\bf H_n$ be a Hadamard matrix of order $n$, then $\bf H_2\otimes H_n$ is a Hadamard matrix of order $2n$. Using this fact, we can enlarge a signature matrix by a factor $2^k$ where $k=1,2,\dots$ and construct Hadamard matrices from only $H_2$.
It is important to mention that although the differences between capacity of different criteria may not be noticeable, for large scale matrices it becomes important.

The enlarged matrices can be decoded using a new simple ML decoder with significant reduction in the complexity of ML decoding. This decoder was originally introduced by \cite{PedramIT} for binary matrices and its none binary version discussed in \cite{Enigma} and in this paper we generalize it to real matrices.

\subsection{The new ML decoder Algorithm}\label{subsec:Mldecoding}

The following is the step by step procedure of this decoder:\\
\begin{enumerate}
\item Define $Z=\sqrt{k}H_k^{-1}\otimes I_m Y$ and split it into $Z^i_{m \times 1}, i=1,2,\dots,k$.
\item Decode $Z^i$'s using ``tensor decoder" \cite{PedramIT} and obtain $X^i$'s.
\item Join $X^i$'s to construct $X$ such that $X=[{X^i}^T, {X^i}^T,\dots,{X^k}^T]$.\\
\end{enumerate}
The following is an example the Following example clarifies this algorithm:\\

\newtheorem{example}{\bf{Example}}
\begin{example}
Suppose that the enlarged signature matrix is ${\bf D}=\frac{1}{\sqrt{2}}{\bf H}_{2}\otimes {\bf A_{4\times 5}}$, where ${\bf A_{4\times 5}}$ is $\bf A_4$ from Table~\ref{tab:GA_CAP} in the appendix; consequently,
\begin{equation*}
{\bf D}=\frac{1}{\sqrt{2}}
\begin{bmatrix}
{\bf A} & {\bf A}\\
{\bf A} & -{\bf A}
\end{bmatrix}
\end{equation*}
Assume that $X=[1,1,-1,-1,-1,-1,-1,1,1,-1]^T$, then
\begin{eqnarray*}
Y={\bf D}X+N=
\begin{bmatrix}
-1.4586\\
-0.5227\\
-0.8251\\
-1.3148\\
0.9584\\
-0.1522\\
3.7170\\
2.0180
\end{bmatrix}
\end{eqnarray*}
where $N$ is an $8\times 1$ AWGN vector. For decoding at the receiver, we split $Y$ into two equal length vectors $Y^1$ and $Y^2$ such that $Y=[Y^1,Y^2]^T$. We then have,
\begin{eqnarray*}
&Y^1=&[-1.4586,-0.5227,-0.8251,-1.3148]^T\\
&Y^2=&[0.9584,-0.1522,3.7170,2.0180]^T\\
\end{eqnarray*}\
Define:
\begin{eqnarray*}
Z^1=&\frac{1}{\sqrt{2}}[Y^1+Y^2]\\
Z^2=&\frac{1}{\sqrt{2}}[Y^1-Y^2]
\end{eqnarray*}
The decoding of $Z^1,Z^2$ using the tensor decoder yields $X^1=[1,1,-1,-1,-1]^T$ and $X^2=[-1,-1,1,1,-1]^T$ where $X=[X^1,X^2]^T$.
\end{example}

In general, for an $km\times kn$ signature matrix which is enlarged from an $m \times n$ matrix, the usual ML decoding needs $2^{km\times kn}$ Euclidean distance measurements while the new decoding needs $k2^{n-m}$ Euclidean distance measurements.
As an example, suppose that we have a $64\times 80$ signature matrix which is enlarged from a $4\times 5$ signature matrix. The ML decoding for this matrix needs $2^{64}*2^{80}=2^{142}$ Euclidean distance computations while using the ``tensor ML decoder", we only need $16*2=32$ Euclidean distance measurements.

\section{Conclusion and Future Work}\label{sec:conclusion}
In this paper, we have defined and derived the capacity for a specific signature matrix and a given $E_B/N_0$ for overloaded CDMA systems. In order to find sub-optimum matrices, a number of optimization criteria have been introduced; namely, the capacity, BER and the distance criteria. We have modelled these criteria and shown that they differ with each other in terms of optimality and complexity. Our simulation results demonstrate that our proposed ED criterion presents the best performance regarding the complexity and optimality accuracy. To derive the sub-optimum signature matrices, we applied the GA as well as the PSO algorithms.
In addition to the real valued matrices, we have also applied our methods in finding sub-optimum binary signature matrices.
For large scale systems, instead of directly optimizing signature matrices, we enlarge small sub-optimal matrices using Kronecker products. We have shown that the capacity of these enlarged matrices is increased by the enlargement factor $k$. We can employ simple ML decoding for such enlarged matrices which significantly reduces the implementation complexity while maintaining optimality.
While in this work, we have applied the GA and the PSO algorithms for optimization purposes,in order to avoid local minima, we suggest to work on ``simulated annealing" as future works. Furthermore, we suggest derivation of sub-optimum matrices for non binary discrete valued signatures.

\section*{Acknowledgements}
The authors would like to thank A. Rashidinejad, M. H. Lotfi Froushani and P. Pad for their contributions and useful discussions.

\bibliographystyle{IEEEtran}
\bibliography{paper_ieee}

\appendix
\begin{table*}[h]
\caption{Sub-optimum matrices based on the capacity method using GA,(note: $A_2$ is optimized for $E_b/N_0=11dB$ and other matrices are optimum for $E_b/N_0=8dB$).}
\begin{tabular}{p{1cm}|cc}
\multicolumn{2}{c}{\large{Capacity~method}}\\ \hline
 2 by 5 &$A_1=\begin{bmatrix}
0.1235 &   0.3177    &0.7605    &0.8739  &  0.4069 \\
0.3723  &  0.9240    &0.5021   & 0.0154  & -0.2553
\end{bmatrix}$
&
 $A_2=\begin{bmatrix}
 -0.3724   & 0.7299 &  -0.0115 &   1.0000 & 0.5408\\
 -0.5254 &   0.2538&   0.9584 &  -0.3117  & -0.7224
 \end{bmatrix}$
 \\
 3 by 4 &
 $A_3=\begin{bmatrix}
 0.9764&    0.3895&    0.7448&   -0.9375\\ -1.0000&    0.1711&    0.4241&	  0.6451\\
 0.8529&    0.6424&    0.0930&    1.0000
 \end{bmatrix}$
 & 
\\
4 by 5& 
$A_4=\begin{bmatrix}
1 & 1 & 0.969 & 0.468 &1 \\
0.424 & -1 & 0.5 & -0.871 &0.5\\
1 & 0.015 & -0.906 &-0.75 &0.719\\
0.430 & 0.995 & -0.938 & 0.984 & 0.984
\end{bmatrix}$
&
 $A_5=\begin{bmatrix}
 1&  1&  1&  1&  1\\
 1& -1&  1& -1&  1\\
 1&  1& -1& -1&  1\\
 1& -1& -1&  1& -1\\
 \end{bmatrix}$
\end{tabular}
\label{tab:GA_CAP}
\end{table*}

\begin{table*}[h]
\caption{Sub-optimum matrices based on the ED method using GA,(the matrices are optimum for $E_b/N_0=8dB$).}
\label{tab:GA_ED}
\begin{tabular}{lll}
\large{ED~method}\\ \hline
$A_1=\begin{bmatrix}
 	0.0591&    0.8787&   -0.6226&    0.4163&    0.2166\\
 	-0.9198&    0.1760&   0.1907 &   0.6094&    0.8851
\end{bmatrix}$
&
$A_2=\begin{bmatrix}
	0.9572&    0.4704&    0.5922&    0.1288\\
	-1.0000&   0.8393&    0.3621&    0.7090\\
	 0.3995&   0.6776&   -0.7468&   -0.1777
\end{bmatrix}$
&
$A_3=\begin{bmatrix}
	-1&  -1&  1&  -1&  1\\ 
	-1&   1& -1&   1&  1\\  
	 1&   1& -1&  -1&  1\\  
    -1&  -1& -1&  -1& -1\\
\end{bmatrix}$
\end{tabular}
\end{table*}
\begin{table*}[h]
\caption{Sub-optimum matrices based on the QD, MD, and BER methods using GA,(the matrices are optimum for $E_b/N_0=8dB$).}
\begin{tabular}{p{10mm}|ccc}
\multicolumn{1}{c}{Size}& QD~method& MD method& BER method\\ \hline
3 by 4&
$\begin{bmatrix}
	0.4520&   -0.3740&    0.9029&    0.1059\\
   -0.7780&    0.3048&    0.9585&   -0.6561\\
    0.9163&    0.4018&    0.3265&   -0.0717
	\end{bmatrix}$
&	
$\begin{bmatrix}
	0.5924&    0.1238&   0.4630&   -0.4371\\
	0.0557&    0.4388&   0.6436&	0.5020\\
	0.9595&    0.4137&   0.0075&    0.4325
	\end{bmatrix}$
&
$\begin{bmatrix}
0.2502&    0.4917&    0.1048&   -0.9300\\
0.6206&    0.9009&   -0.9958&    0.4022\\
0.9903&    0.2592&    0.4383&	 0.9961
\end{bmatrix}$
\\
2 by 5&
$\begin{bmatrix}
	 0.5315&    0.9989&   -0.9456&    0.5273&    0.4257\\
	 0.4364&    0.3203&    0.5859&   -0.9514&    0.7039
\end{bmatrix}$
& &
\end{tabular}
\label{tab:GA_Other}
\end{table*}

\begin{table*}[h]
\caption{Sub-optimum matrices based on various methods using PSO,(the matrices are optimum for $E_b/N_0=8dB$).}
\begin{tabular}{l|cc}
\multicolumn{1}{c}{Mthod}& 2 by 5& 3 by 4\\ \hline
Capacity &
$\begin{bmatrix}
1.0000&        0&    1.0000&    1.0000&   -0.3120\\
0.9419&    1.0000&  -0.6067&    0.0812&    0.6859
\end{bmatrix}$
&
$\begin{bmatrix}
	 0&         0&    1.0000&    0.3137\\
	 0&    1.0000&    1.0000&         0\\
1.0000&         0&    1.0000&	      0
\end{bmatrix}$
\\\hline
ED&
$\begin{bmatrix}
	 1.0000&    1.0000&    1.0000&    0.5432&   -0.0269\\
	 0.5206&    0.1099&   -0.2031&    1.0000&    1.0000
\end{bmatrix}$
&
$\begin{bmatrix}
	1.0000&         0&    0.0665&    1.0000\\
	     0&    1.0000&         0&    1.0000\\
	     0&         0&    1.0000&	1.0000
\end{bmatrix}$
\\\hline
MD&
$\begin{bmatrix}
	0.3045&    0.6719&    1.0000&    0.2925&   -0.0804\\
    1.0000&    0.2708&    0.0711&   -0.7045&    1.0000
\end{bmatrix}$
&
$\begin{bmatrix}
	1.0000&         0&    1.0000&    0.0483\\
	1.0000&    1.0000&    0.0574&         0\\
	1.0000&    0.0701&         0&	  	1.0000
\end{bmatrix}$
\\\hline
BER&
$\begin{bmatrix}
	1.0000&   -0.7644&         0&    1.0000&    0.4113\\
	1.0000&    1.0000&    0.5402&   -0.4707&    1.0000
\end{bmatrix}$
&
$\begin{bmatrix}
	1.0000&    1.0000&   -0.2516&   -0.9898\\
	1.0000&   -0.1536&    1.0000&   -0.0209\\
	1.0000&         0&   -0.6976&	1.0000
\end{bmatrix}$
\end{tabular}
\label{tab:PSO_various}
\end{table*}
\end{document}